# Defect-related states in MAPbI$_3$ halide perovskite single crystals revealed by the photoluminescence excitation spectroscopy

*Aleksei O. Murzin*, Nikita I. Selivanov, Vadim O. Kozlov, Ivan I. Ryzhov, Alexei V. Emeline, Yury V. Kapitonov**

Saint Petersburg State University, 7/9 Universitetskaya Emb., 199034, St. Petersburg, Russia
*E-mail: alekseymurzin10@gmail.com
** E-mail: yury.kapitonov@spbu.ru

ABSTRACT: The MAPbI$_3$ halide perovskite single crystals are studied at 5 K temperature using the photoluminescence excitation spectroscopy. Two non-interacting types of states are determined: bound excitons and defect-related states. Excitation of the crystal with light energy below the bound exciton resonance reveals the complex low-density defects emission, otherwise hidden by the emission of bound excitons. A way to separate these defect-related luminescence spectra is proposed, and the thorough study of this emission regime is carried out. The results of this study opens an area of low-density defects and dopants exploration in halide perovskite semiconductors.

Halide perovskite semiconductors are novel optical materials, which have found their perspective applications as an active medium in many optoelectronic devices: solar cells,[1] light emitting diodes,[2] micro-lasers,[3-5] electromagnetic radiation detectors[6] and others. In all these applications defect states, which energy levels located within the band-gap of material, plays a crucial role being either an undesired obstacle, or a mean of the material functionalization. The rich information about the energy states and transitions between them could be obtained by optical methods including widely used photoluminescence (PL) spectroscopy. Since the light emission occurs from the lowest energy states, this method is able to reveal the low-density but still important defect and dopant states in the material band-gap.

Optical properties of direct band-gap halide perovskites are determined to a large extent by the exciton resonance in these materials.[7-9] The exciton binding energy of tens meV in the most studied 3D lead-halide perovskites APbX$_3$ (where A = Cs, MA (CH$_3$NH$_3$), X = I, Br, Cl)[9] makes these resonances being important even in polycrystalline films at room temperature.[10] However, the most fundamental results can be obtained by studying the optical properties of their single crystals at temperatures close to 0 K.[7,8,11,12] The most general pattern observed in such experiments for abovementioned materials is the free exciton (FE) state being responsible for the resonance in the reflectivity, and a sharp Stokes-shifted for several meV bound exciton (BE) state, followed by the long low-energy tail.[12] Typically this tail has poorly resolved or no structure, and therefore, it provides not much information about defect-related states in the material. Nevertheless, this tail states play an important role in lasing phenomenon, as it was demonstrated that the laser gain spectral region is located well below the BE resonance.[5]

In this study we demonstrate that the PL from defect-related states can be revealed by the careful and precise selection of the excitation energy ($E_{ex}$). We use the micro-photoluminescence excitation (PLE) spectroscopy to study the MAPbI$_3$ single crystals at 5 K temperature. At the excitation above the BE resonance the conventional PL spectrum is observed. Below the energy of BE resonance the absorption rapidly decreases, and the crystal becomes effectively "transparent". Consequently, the emission from defect-related states could be clearly seen in the PL spectra demonstrating their complex and informative energy structure.

MAPbI$_3$ (MA=CH$_3$NH$_3$) single crystals of several millimeters in size were grown from solution. Single perovskite phase was confirmed by the XRD measurements of the grounded crystal (**Figure S1**). Optical behavior of the single crystals was characterized by the micro-reflection and micro-PL spectroscopy at 5 K temperature. **Figure 1 (a)** shows the reflectivity spectrum dominated by the FE resonance at $E_{FE}$ = 1.637 eV. Same Figure shows PL spectra with the band-to-band continuous excitation at $E_{ex}$ = 1.671 eV well above the FE resonance for different excitation intensities $I_{ex}$ (spectra are normalized on $I_{ex}$). At high excitation intensities the sharp BE resonance is observed at $E_{BE}$ = 1.632 eV with the 5 meV Stokes shift from the energy of FE transition. It is followed by the lower energy tail without any noticeable structure. The BE intensity scales superlinearly with the excitation intensity ($\sim I_{ex}^{1.4}$), that provides an additional prove of its excitonic origin. Excitation intensity in further experiments was chosen to $I_{ex}$ = 116 W cm$^{-2}$.

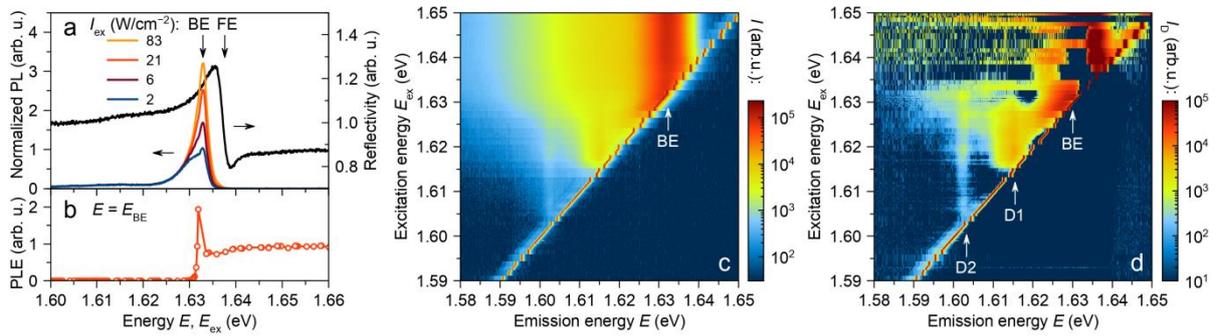

**Figure 1.** (a) Reflectivity spectrum (black) and PL spectra for different excitation intensities $I_{ex}$ normalized on $I_{ex}$. (b) PLE spectrum for BE resonance. (c) 2D PLE spectrum for cross-polarized excitation. (d) 2D PLE spectrum of defect-related states extracted from the data in (c) using Equation 5 for the reference point $E_0$ = 1.580 eV, $E_{ex0}$ = 1.655 eV. All spectra are measured at 5 K.

PL spectra were collected for different excitation energies $E_{ex}$, and the 2D PLE spectrum $I(E, E_{ex})$ was obtained, where $E$ is the emission energy. Such measurements were repeated for the co- and cross-circularly-polarized excitation and detection (**Figure S2 (a,b)**). Except the hot luminescence above the BE resonance the signal demonstrates low polarization degree (below 10%) (Figure S2 (c)). For this reason, the cross-polarized PLE spectrum will be used for the further discussion of spectral properties, since it has reduced amount of the scattered excitation light.

Figure 1 (c) shows the 2D PLE spectrum for the cross-polarized geometry at 5 K. Its cross-section for the BE emission energy ($E = E_{BE}$) is presented in Figure 1 (b). The behavior of the BE PLE spectrum corresponds to the nearly constant absorption above the exciton resonance in conformity with the Elliot formula.[13] However no noticeable increase of the PL intensity was observed when the FE resonance was excited. Similar behavior was reported for MAPbI$_3$ single crystals at low temperature in photocurrent excitation experiments.[11]

PL spectra are remarkably changed at $E_{ex} < E_{BE}$. On the background of the BE-related emission several new narrow emission lines arise at emission energies well below $E_{BE}$. Their narrow line width and large Stokes shift are indicative for emission from deep defect-related states.

The observed step-like behavior of PLE spectra leads us to the conclusion that two non-interacting types of states can be distinguished: the BE exciton, which has very strong absorption with the emission spectrum consisting of a sharp peak with a long structureless tail; and low-density defect-related states which could be observed only at excitation tuned below the BE absorption energy, when the crystal become transparent enough to excite these states. **Figure 2 (a,b)** summarizes this conclusion in the energy diagram of states.

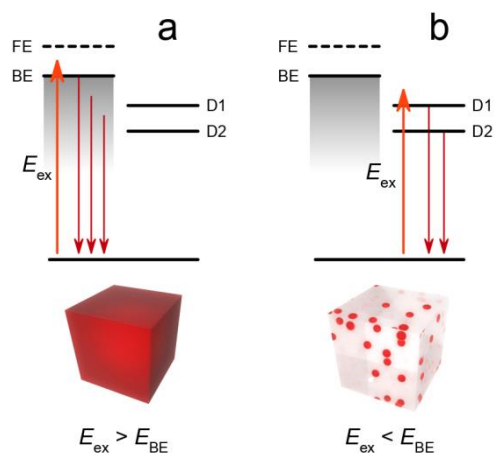

**Figure 2.** Energy diagrams of transitions between states at $E_{ex} > E_{BE}$ (a) and $E_{ex} < E_{BE}$ (b). FE – free exciton, BE – bound exciton, D1, D2 – defect-related states. Below are artist's impressions of the perovskite crystals: non-transparent emitting as a whole from the surface at $E_{ex} > E_{BE}$ (a), and transparent with emitting defect centers in the volume at $E_{ex} < E_{BE}$ (b).

Our assumption can be used to determine spectral properties of defect-related states basing on the experimental PLE spectrum. The total PL emission intensity $I(E, E_{ex})$ at $E \leq E_{ex}$ could be decomposed into two components corresponding to the emission from BE (first component) and the emission from the defect-related state (second component):

$$I(E, E_{ex}) = A(E_{ex}) \cdot S(E) + \big(1 - A(E_{ex})\big) \cdot A_D(E_{ex}) \cdot S_D(E), \quad (1)$$

where $E$ is the emission energy, $E_{ex}$ is the excitation energy, $A(E_{ex})$ and $A_D(E_{ex})$ are absorptance spectra and $S(E)$ and $S_D(E)$ are emission spectra of the BE and the defect-related state respectively. We assume that the defect-related state could be excited only by the light that is not absorbed by the BE and therefore, there is no channel for the relaxation between these two states.

We also assume that at $E_{ex0} \geq E_{BE}$ all light is absorbed by BE ($A = 1$), and the pure BE emission spectrum can be measured:

$$I(E, E_{ex0}) = S(E). \quad (2)$$

Let us assume that the defect-related emission is spectrally localized around the resonance energy $E_D$, while the BE emission tail is spectrally broad. In this case at $E_0 \ll E_D$ the defect-related emission vanishes and the BE absorption could be measured:

$$I(E_0, E_{ex}) = A(E_{ex}) \cdot S(E_0) = A(E_{ex}) \cdot I(E_0, E_{ex0}). \quad (3)$$

Combining Equation 2 and 3 into Equation 1 we obtain:

$$I(E, E_{ex}) = \frac{I(E_0, E_{ex}) \cdot I(E, E_{ex0})}{I(E_0, E_{ex0})} + \frac{I(E_0, E_{ex0}) - I(E_0, E_{ex})}{I(E_0, E_{ex0})} \cdot A_D(E_{ex}) \cdot S_D(E). \quad (4)$$

This equation could be rewritten into the following form:

$$I_D(E, E_{ex}) = A_D(E_{ex}) \cdot S_D(E) = \frac{I(E, E_{ex}) \cdot I(E_0, E_{ex0}) - I(E_0, E_{ex}) \cdot I(E, E_{ex0})}{I(E_0, E_{ex0}) - I(E_0, E_{ex})}, \quad (5)$$

where the left part stand for the pure PLE spectrum $I_D(E, E_{ex})$ of the defect-related state and the rightmost part is composed of the PLE spectral data measured in the experiment. Figure 1 (d) shows the transformation by the Equation 5 of the data from Figure 1 (c) for the reference point $E_0 = 1.580$ eV, $E_{ex0} = 1.655$ eV. This transformation attenuates broad structureless BE emission background and highlights two defect-related resonances D1 at $E_{D1} = 1.615$ eV and D2 at $E_{D2} = 1.603$ eV. The resonantly excited PL from the BE state also can be seen in Figure 1 (d). **Figure 3** (a) and (b) show normalized PL and PLE spectra of these two defect states together with the BE state.

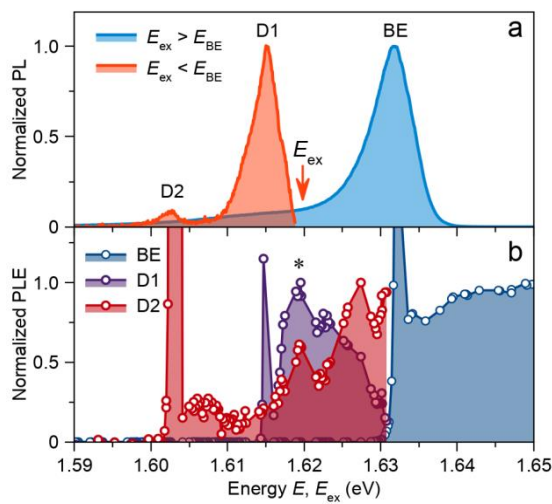

**Figure 3.** (a) Normalized PL spectra with excitation above the BE resonance (blue curve, $E_{ex}$ = 1.665 eV) and below the BE resonance processed using Equation 5 (red curve, $E_{ex}$ = 1.620 eV, denoted by an arrow). BE – bound exciton, D1, D2 – defect-related states. (b) Normalized PLE spectra for the BE (blue curve), D1 (violet curve) and D2 (red curve) resonances. PLE for defect-related states are processed using Equation 5. Absorption maximum of defect-related states is denoted by *.

Figure 3 (a) clearly demonstrates the benefits of the proposed excitation scheme: in the case of the excitation energy above the BE resonance there is no hint of D1 and D2 states (blue curve). However when the excitation is spectrally located below BE and the proposed processing is applied then these two defect-related states become visible (red curve). This processing also allows one to track the PLE of these states shown in Figure 3 (b). Both states have simultaneous maximum in absorption denoted by "*" and located at 5 meV above the D1 transition. In contrast to the BE state, the D1 state has no resonant PL. **Figure S3** summarizes proposed transitions between states in the diagram.

The above analysis was performed for one point of the sample, determined by the diameter of the exciting laser beam (~10 μm). Despite the fact that the BE emission was fairly uniform over millimeter-size sample, the defect-related emission was extremely sensitive to the spatial point of excitation. This indicates that observed defects have sufficiently low density. **Figure**

**S4** (a-c) show spectra for several different sample points. Spectral positions of defect-related states in these points lies in the range 1.600-1.630 eV. Still in all of these points the general pattern remains valid: at the excitation with energies below the BE transition the structureless background decreases and sharp defect-related peaks are seen. However, one should take care applying Equation 5, since in some cases defect-related PL could have a component even broader then the BE emission (Figure S4 (d)), and, therefore, Equation 3 is no more valid. Microscopic mechanisms behind observed phenomena is beyond the scope of this study. Despite this, several general conclusions could be drawn based on the obtained data. Defect-related states are spectrally located in the transparency region of the sample. Therefore, the absence of their PL in the case of the excitation tuned to the BE state proves that there is no excitation transfer from BE to defect-related states. For low-density defects this transfer could be limited by the slow exction diffusion (caused by the localization in the BE state), small trapping cross-section of defects, or short exciton lifetime (limited by the radiative lifetime of the BE state). Another puzzle is the origin of the BE emission tail extended far below the defect-related states. One of the possible explanations is the exciton self-trapping mechanism, well-known for low-dimensional halide perovskites,[14] and being able to lower the exciton energy by a great amount.

In conclusion, we studied the $MAPbI_3$ halide perovskite single crystal at 5 K temperature using the photoluminescence excitation spectroscopy. The 2D PLE spectroscopic data clearly indicates the coexistence of two non-interacting types of states in the crystal: bound excitons characterized by the strong absorption with emission consisting of a sharp peak followed by a structureless low-energy tail; and defect-related states manifesting themselves with the sharp emission peaks well below the BE state. Defect-related states could be revealed by the careful selection of the excitation energy below the BE resonance. Our findings provide the basis for further study of low-density defects and doping in halide perovskite single crystals, opening the way for the temperature- and intensity-dependent studies of defect-related states emission, polarization-resolved magnetooptical experiments, photon echo spectroscopy and polarimetry,[15] and other optical spectroscopy methods.

**Experimental Section**

*Materials*: Lead (II) iodide $PbI_2$ (99,999 %, Sigma-Aldrich), hydroiodic acid HI (56% in $H_2O$, Iodobrom), methylamine $CH_3NH_2$ (38% in $H_2O$, Lenreactiv) were used as received. Silica hydrogel was obtained using the sodium metasilicate crystallohydrate ($Na_2SiO_3·9H_2O$) solution. Distilled water was used as a solvent.

*Synthesis and characterization*: $MAPbI_3$ single crystals were grown from solution in the hydrogel of orthosilicic acid by the method analogous to the described in Ref. [16]. Details of the synthesis method will be described elsewhere. The XRD measurements were performed with diffractometer Bruker "D8 Discover" using a long focus X-ray tube CuKα anode. Reflected X-rays were detected using a solid position-sensitive detector LYNXEYE.

*Optical study*: The $MAPbI_3$ single crystal was mounted into the closed-cycle helium cryostat Montana Cryostation and cooled down to 5 K. The sample irradiation and detected luminescence were through 20x microobjective lens in the autocollimation geometry. Spectra were obtained with the spectral resolution 0.2 meV using a spectrometer build up from the MDR-4 monochromator equipped with a CCD-detector. For normal reflectivity measurements a halogen lamp was used. For PL excitation a continuous wave Ti:Sapphire laser Tekhnoscan TD Scan was used. Laser was focused down to a 10 μm-diameter spot on the sample. The linearly-polarized excitation laser beam and PL emission were directed through the λ/4-waveplate. A λ/2-waveplate followed by a thin-film polarizer were installed in front of the entrance slit of the spectrometer. The PL signal with co- or cross-circular-polarization with respect to the excitation laser beam could be obtained by corresponding orientation of the λ/2-waveplate. For PLE measurements the laser wavelength was scanned. The LPC BEOC laser intensity stabilizer was used. For PLE spectra procession the reference spectra were averaged over 4.3 meV (60 PL spectra) over $E$ axis, and over 12 meV (12 PLE spectra) over $E_{ex}$ axis.

**Supporting Information**
Supporting Information is available from the Wiley Online Library or from the author.

**Acknowledgements**
This study was supported by the Russian Science Foundation (project 19-72-10034). This work was carried out on the equipment of SPbU Resource centers "Nanophotonics", "X-ray Diffraction Studies" and partly using research facilities of the laboratory "Photoactive nanocomposite materials" supported within SPbU program (ID: 51124539).
Received: ((will be filled in by the editorial staff))
Revised: ((will be filled in by the editorial staff))
Published online: ((will be filled in by the editorial staff))

References

[1] A. K. Jena, A. Kulkarni, T. Miyasaka, *Chem. Rev*. **2019**, 119 (5), 3036-3103.

[2] L. Chouhan, S. Ghimire, C. Subrahmanyam, T. Miyasaka, V. Biju, *Chem. Soc. Rev*. **2020**.

[3] H. Dong, C. Zhang, X. Liu, J. Yao, Y. S. Zhao, *Chem. Soc. Rev*. **2020**, 49, 951-982.

[4] I. Shishkin, A. Polushkin, E. Tiguntseva, A. O. Murzin, B. V. Stroganov, Yu. V. Kapitonov, S. A. Kulinich, A. Kuchmizhak, S. Makarov, *Appl. Phys. Express* **2019**, 12, 122001.

[5] A. O. Murzin, B. V. Stroganov, C. Günnemann, S. B. Hammouda, A. V. Shurukhina, M. S. Lozhkin, A. V. Emeline, Yu. V. Kapitonov, arXiv:2003.07927 **2020**.

[6] J. Deng, J. Li, Z. Yang, M. Wang, *J. Mater. Chem. C* **2019**, 7, 12415-12440.

[7] J. Tilchin, D. N. Dirin, G. I. Maikov, A. Sashchiuk, M. V. Kovalenko, E. Lifshitz, *ACS Nano* **2016**, 10 (6), 6363-6371.

[8] O. A. Lozhkina, V. I. Yudin, A. A. Murashkina, V. V. Shilovskikh, V. G. Davydov, R. Kevorkyants, A. V. Emeline, Yu. V. Kapitonov, D. W. Bahnemann, *J. Phys. Chem. Lett*. **2018**, 9 (2), 302-305.

[9] Y. Jiang, X. Wang, A. Pan, *Adv. Mater*. **2019**, 31 (47), 1806671.

[10] N. Sestu, M. Cadelano, V. Sarritzu, F. Chen, D. Marongiu, R. Piras, M. Mainas, F. Quochi, M. Saba, A. Mura, G. Bongiovanni, *J. Phys. Chem. Lett*. **2015**, 6 (22), 4566-4572.

[11] L. Q. Phuong, Y. Nakaike, A. Wakamiya, Y. Kanemitsu, *J. Phys. Chem. Lett*. **2016**, 7 (23), 4905-4910. (Note that in this article the BE resonance is attributed to the FE resonance).

[12] V. I. Yudin, M. S. Lozhkin, A. V. Shurukhina, A. V. Emeline, Yu. V. Kapitonov, *J. Phys. Chem. C* **2019**, 123 (34), 21130-21134.

[13] R. J. Elliott, *Phys. Rev.* **1957**, 108 (6), 1384.

[14] S. Li, J. Luo, J. Liu, J. Tang, *J. Phys. Chem. Lett*. **2019**, 10 (8), 1999-2007.

[15] S. V. Poltavtsev, Yu. V. Kapitonov, I. A. Yugova, I. A. Akimov, D. R. Yakovlev, G. Karczewski, M. Wiater, T. Wojtowicz, M. Bayer, *Sci Rep* **2019**, 9, 5666.

[16] N. I. Selivanov, Yu. A. Rozhkova, R. Kevorkyants, A. V. Emeline, D. W. Bahnemann, *Dalton Trans*. **2020**, 49, 4390-4403.

**TOC entry:** The MAPbI$_3$ halide perovskite single crystals are studied at 5 K temperature using the photoluminescence excitation spectroscopy. Excitation of the crystal with light energy below the bound exciton resonance reveals complex low-density defects emission, otherwise hidden by the emission of bound excitons.

**Keyword:** halide perovskites

A. O. Murzin[*], N. I. Selivanov, I. I. Ryzhov, V. O. Kozlov, A. V. Emeline, Yu. V. Kapitonov[**]

**Title** Defect-related states in MAPbI$_3$ halide perovskite single crystals revealed by the photoluminescence excitation spectroscopy

**ToC figure:**

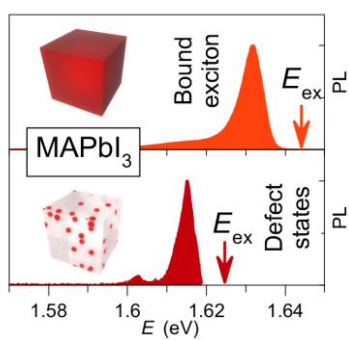

# Supporting Information

**Title** Defect-related states in MAPbI$_3$ halide perovskite single crystals revealed by the photoluminescence excitation spectroscopy

Aleksei O. Murzin*, Nikita I. Selivanov, Ivan I. Ryzhov, Vadim O. Kozlov, Alexei V. Emeline, Yury V. Kapitonov**

Saint Petersburg State University, 7/9 Universitetskaya Emb., 199034, St. Petersburg, Russia
*E-mail: alekseymurzin10@gmail.com
** E-mail: yury.kapitonov@spbu.ru

SUPPORTING INFORMATION

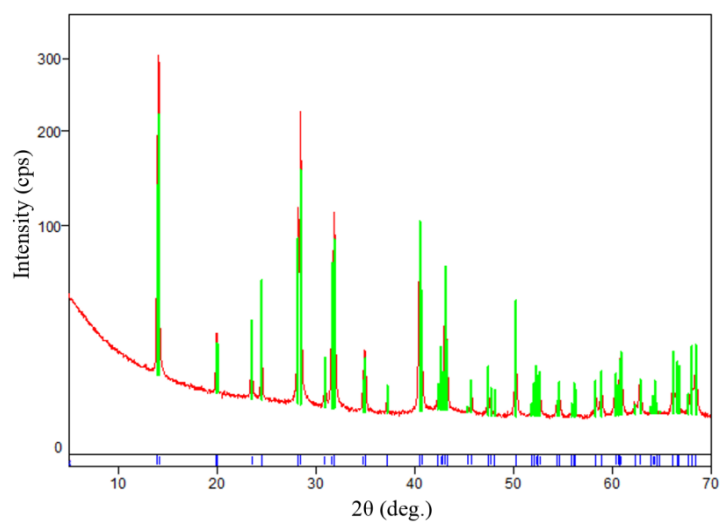

**Figure S1.** XRD spectrum from the powdered MAPbI$_3$ single crystal. Red – experimental data; blue – database record Ref. number 01-083-7582, green – fit.

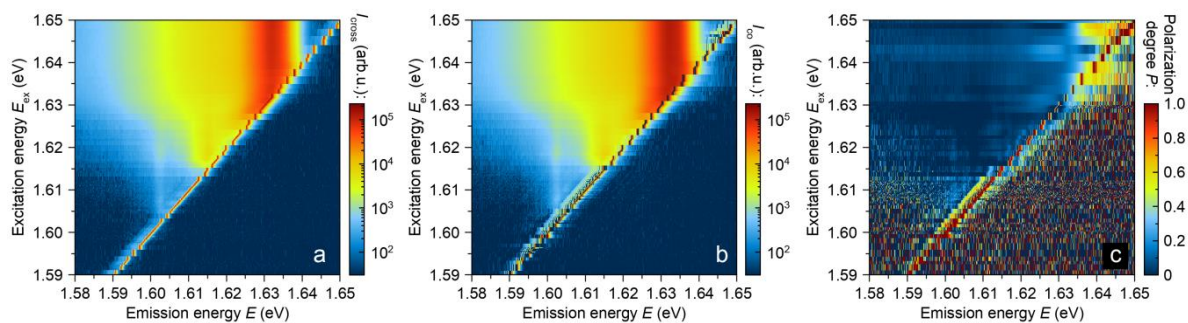

**Figure S2.** PLE spectra for the cross-polarized $I_{cross}$ (a) and co-polarized $I_{co}$ (b) detection and excitation polarizations. PL polarization degree (c) defined as $P = (I_{co} - I_{cross})/(I_{co} + I_{cross})$.

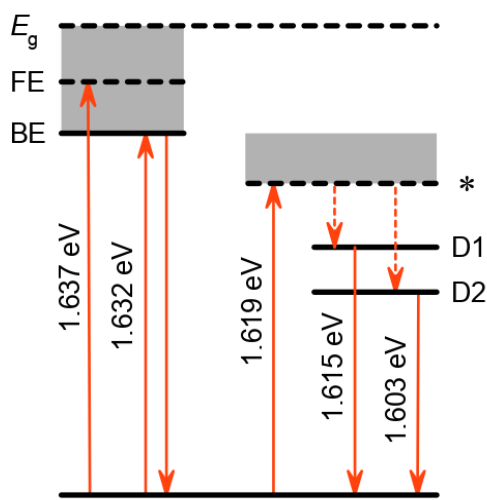

**Figure S3.** Proposed diagram of transitions between states at 5 K for the sample point discussed in the main text. $E_g$ – band gap, FE – free exciton, BE – bound exciton, D1 and D2 – defect related states, * – absorption maximum of defect-related states. Solid arrows denote light absorption or emission transitions, dashed arrows – non-radiative relaxation. States shown dashed could be seen only in absorption or reflection measurements but not in emission.

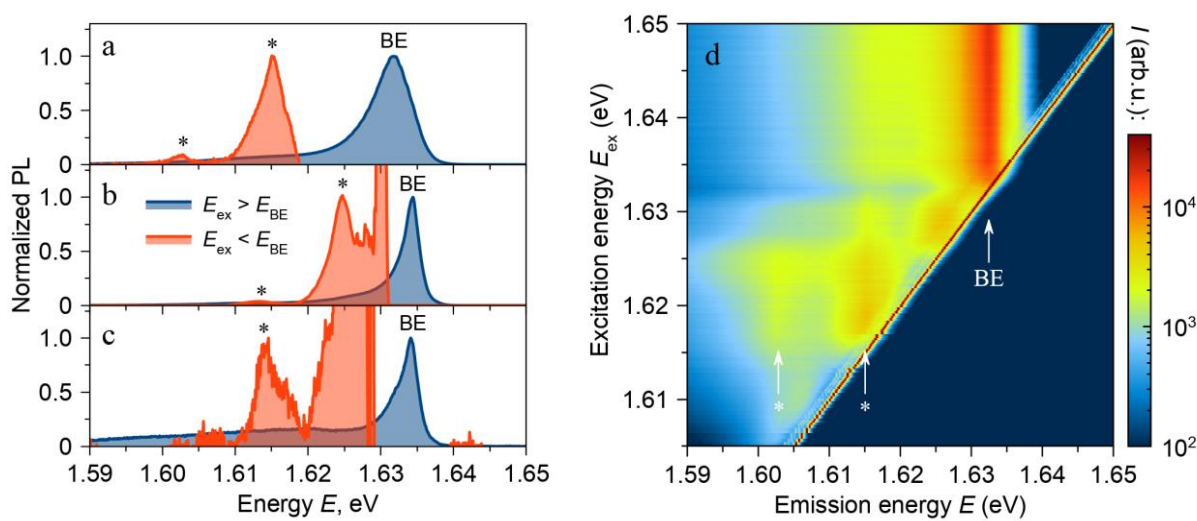

**Figure S4.** (a-c) Normalized PL spectra with excitation above the BE resonance (blue curves) and below the BE resonance processed using Equation 5 (red curves) for different sample points. BE – bound exciton, * – defect-related states. (d) 2D PLE spectrum for the sample point, where processing by Equation 5 could not be made due to the broad component in defects emission. However several defect-related states (*) could be resolved.